\documentclass[aps,prl,twocolumn,superscriptaddress,preprintnumbers]{revtex4-2} 

\usepackage{amsmath,amssymb,amsfonts} 
\usepackage{graphicx} 
\usepackage{hyperref}
\usepackage{bm} 
\usepackage{times} 
\usepackage{xcolor}
\usepackage{physics}
\usepackage{multirow}
\usepackage{pifont}

\definecolor{darkgreen}{HTML}{006400}

\begin{document}

\title{Resource complexity of Symmetry Protected Topological phases}

\author{Alberto Giuseppe Catalano}
\affiliation{Dipartimento di Fisica e Astronomia "G. Galilei" \& Padua Quantum Technologies
Research Center, Università degli Studi di Padova, Italy I-35131, Padova, Italy}
\affiliation{
INFN, Sezione di Padova, via Marzolo 8, I-35131, Padova, Italy}

\author{Sven Benjamin Kožić}
\affiliation{Institut Ruder Bošković, Bijenička cesta 54, Zagreb 10000, Croatia}

\author{Gianpaolo Torre}
\affiliation{Institut Ruder Bošković, Bijenička cesta 54, Zagreb 10000, Croatia}

\author{Carola Ciaramelletti}
\affiliation{Dipartimento di Ingegneria e Scienze dell’Informazione e Matematica, Università dell’Aquila, via Vetoio, I-67010 Coppito-L’Aquila, Italy.}

\author{Simone Paganelli}
\affiliation{Dipartimento di Scienze Fisiche e Chimiche, Università dell’Aquila, via Vetoio, I-67010 Coppito-L’Aquila, Italy.}

\author{Fabio Franchini}
\affiliation{Institut Ruder Bošković, Bijenička cesta 54, Zagreb 10000, Croatia}

\author{Salvatore Marco Giampaolo}
\affiliation{Institut Ruder Bošković, Bijenička cesta 54, Zagreb 10000, Croatia}

\begin{abstract}
We pursue the identification of quantum resources carried by topological order,
by evaluating quantum magic, quantified through the rank-$2$ Stabilizer R\'enyi entropy $\mathcal{M}_2$, in one-dimensional systems hosting symmetry-protected topological phases (SPTP). 
Focusing on models with an exact duality between an SPTP and a trivial one, namely the dimerized XX and the Cluster–Ising chains, we show that dual points exhibit identical amounts of magic, even though they belong to distinct topological sectors. 
A subextensive asymmetry arises only under open boundary conditions, where edge effects break the duality, but this correction is non-topological and depends on microscopic parameters. 
These results stand in contrast to the case of topological frustration, where delocalized excitations enhance the magic logarithmically with system size. 
They also complement recent analyses in the literature, showing that the total magic is largely insensitive to the presence of topological order, hence suggesting that topological order is not necessarily a genuine computational resource.

\end{abstract}

\preprint{RBI-ThPhys-2025-38}
\maketitle
A key challenge in quantum information science is to identify the resources that a quantum state must possess to perform tasks beyond the reach of classical computation, that is, to establish quantum supremacy~\cite{Harrow2017}. 
This question naturally connects to the study of quantum many-body systems, whose ground states often embody such resources.
For a long time, entanglement~\cite{Jozsa2003, Horodecki2009} was believed to be the main bridge between many-body physics and quantum information~\cite{amico2008, Latorre_2009, Eisert2010, DeChiara2012, Laflorencie2016}. 
However, it has become clear that entanglement alone cannot fully account for quantum advantage. 
Some highly entangled states can be efficiently simulated using classical resources~\cite{Terhal2002}, while certain computational models exhibit quantum advantage even with little entanglement~\cite{Knill1998}. 
This insight has motivated a broader search for other operationally meaningful quantum resources~\cite{Gour_2025}.
Among these, quantum magic (or nonstabilizerness)~\cite{Howard2014, Seddon2019, Veitch2014, Anders2009} has recently emerged as a key quantity, measuring how far a quantum state lies outside the stabilizer manifold. 
The latter is the set of states obtainable from computational-basis states through Clifford circuits~\cite{Aaronson2004, Gottesman1998, Bravyi2018}, which themselves can be efficiently simulated on classical computers. 
Therefore, in loose terms, magic can be interpreted as the amount of non-Clifford resources needed to prepare a given quantum state starting from a computational basis state. 
It is thus unsurprising that, similarly to entanglement in the past, magic has become a powerful proxy linking phases of matter to their computational capabilities~\cite{Oliviero2022, Odavic2023, Catalano2025}.

Among quantum phases, topological ones are especially relevant, as they provide intrinsic protection against local perturbations and encode information nonlocally~\cite{Nayak2008}. 
They exhibit correlations beyond local order parameters, host long-range entanglement, and enable fault-tolerant computation through the braiding of anyonic excitations~\cite{Kitaev2003, Freedman2003}. 
Given these properties, one might expect topologically ordered phases to exhibit a higher complexity and thus host a large quantum magic, reflecting their highly nonclassical structure. 
However, this connection is not universal: the paradigmatic toric code~\cite{Kitaev2003}, though topologically ordered, admits as ground state a stabilizer state with no magic~\cite{Gottesman1998, Bravyi2018}.

In one-dimensional systems, the topological order is more fragile.
Still, a weaker analog exists in the form of symmetry-protected topological phases (SPTPs)~\cite{Chen2011, Pollmann2012}, that are phases that present topological properties only as long as symmetry-breaking terms are excluded. 
Examples include the one-dimensional cluster-Ising model~\cite{Son2011, Smacchia2011, Giampaolo2014, Maric2020a} and the dimerized XX chain~\cite{Campos2007, Bahovadinov2019}, which can be obtained from the Su–Schrieffer–Heeger (SSH) model~\cite{Su1979, Batra2020, Ciaramelletti2025} via a Jordan–Wigner transformation.
Compared to inherent topological phases, topological entanglement entropy~\cite{Kitaev2006, Levin2006} is not uniquely defined in SPTPs. 
For instance, under open boundary conditions, the emergence of edge states gives rise to nonvanishing edge-to-edge entanglement~\cite{Franchini2014, Micallo2020}.
Similarly, only the weaker form of symmetry-preserved long-range entanglement applies, and a finite-depth unitary circuit can remove all entanglement if the symmetry is not enforced, whereas this is not possible for true topological phases~\cite{Chen2010, Chen2011, Bravyi2006}.
The distinction between these two categories becomes less sharp when shallow circuits are supplemented by measurement-based protocols, as it appears that certain types of inherent topological order phases can be created with finite-depth circuits~\cite{Piroli2021, Piroli2024}.
Nevertheless, SPT phases present relevant properties, even from a computational point of view. 
In particular, one-dimensional SPT ground states (e.g., the cluster state) equip quantum wires for measurement-based quantum computation (MBQC)~\cite{Else2012, Miller2015, Ellison2021}, while universal MBQC requires higher-dimensional resource states~\cite{Raussendorf2001, Raussendorf2006, Bartlett2006, Briegel2009, Stephen2017, Raussendorf2017, Adhikary2024, Li2025}.

In this Letter, we explore the connection between computational complexity and topological phases by evaluating the quantum magic of one-dimensional SPTPs and comparing it with that of their topologically trivial counterparts. 
To make this comparison stringent, we focus on models possessing an exact duality between an SPT and a trivial phase, namely the aforementioned dimerized XX and Cluster–Ising chains. 
We find that dual points exhibit identical amounts of magic whenever the duality is preserved (e.g., under periodic boundary conditions), whereas a subextensive asymmetry arises only with open boundaries, where edge effects explicitly break duality. 
However, this correction is not of topological origin, as it depends on microscopic Hamiltonian parameters, and, as we will show in the following, a similar phenomenology also occurs in the quantum Ising chain. 
Our findings, therefore, suggest that assessing computational complexity may require considering a different quantity, such as the long-range magic (LRM)~\cite{Ellison2021, korbany2025, wei2025}, which captures the component of nonstabilizerness that cannot be eliminated by shallow local circuits.

\begin{figure}[t!]
\centering
\includegraphics[width=0.95\columnwidth]{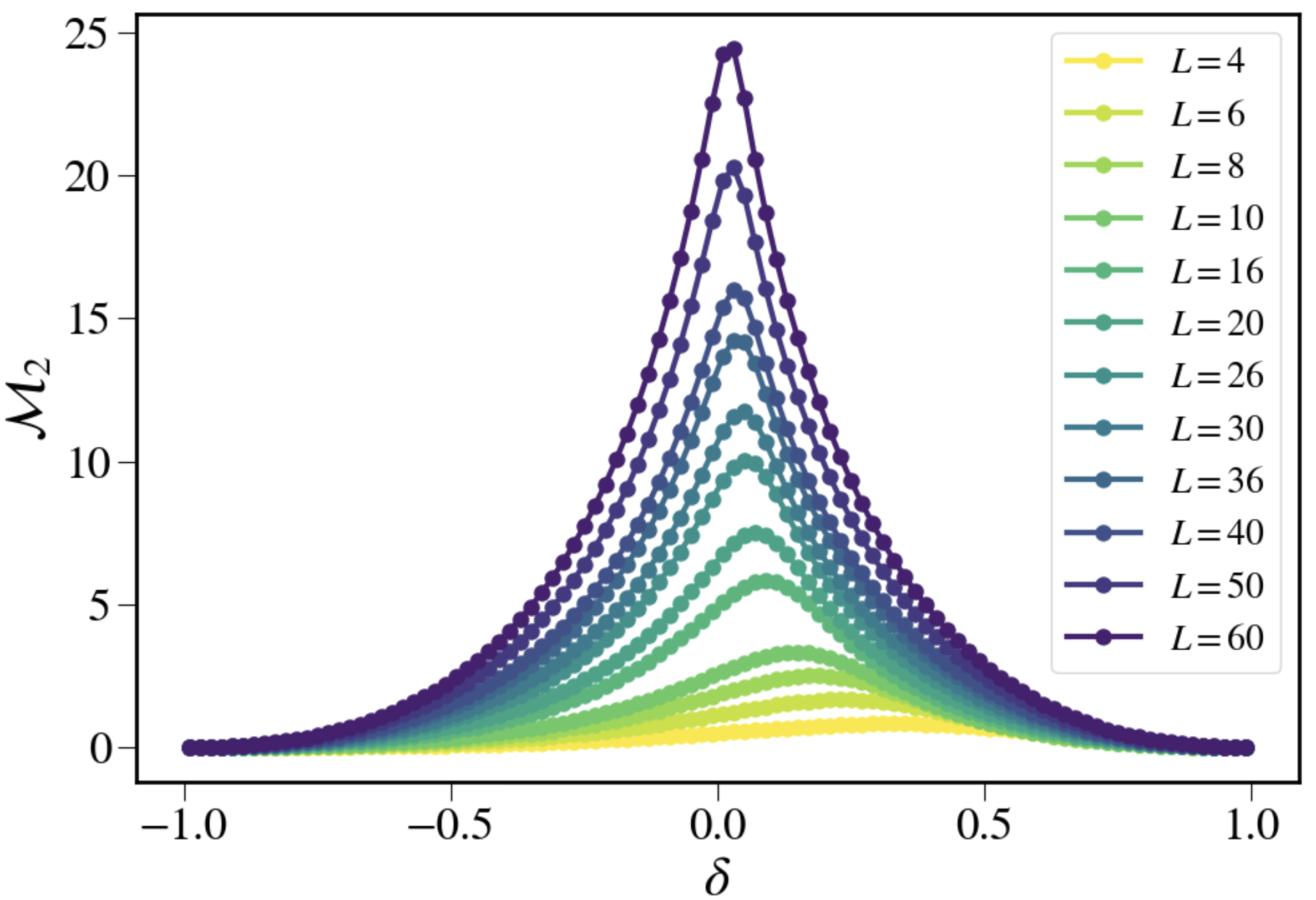}
\caption{
Behavior of $\mathcal{M}_2$ for the SSH/dimerized XX model in Eq.~\eqref{eqn:Hamiltonian_SSH} as a function of the dimerization parameter $\delta$ for different system sizes under OBCs. 
The maximum lies in the SPTP and shifts toward the critical point $\delta = 0$ as the chain length increases.
}
\label{Magic_SSH}
\end{figure}

Before proceeding with our analysis, we note that, throughout this letter, magic is quantified using the stabilizer Rényi entropies (SREs)~\cite{Leone2022}. 
For a pure state $\ket{\psi}$ of $L$ qubits, the SRE of order $\alpha$ is defined as
\begin{equation}
\label{eq:SRE_Definitiom}
 \mathcal{M}_\alpha (\ket{\psi} )=\frac{1}{1-\alpha}\log_{2}{ \left( \frac{1}{2^{L}}\sum_{\mathcal{P}} |\bra{\psi} \mathcal{P} \ket{\psi}|^{2\alpha} \right)}. 
\end{equation}
Here, $\mathcal{P}=\bigotimes_{i=1}^L P_i$ runs over all possible Pauli strings, $P_{i} \in \{ \sigma^0_{i}, \sigma_{i}^{x}, \sigma_{i}^{y}, \sigma_{i}^{z} \}$ with $\sigma_i^{0}$ representing the identity operator, and the order $\alpha$ is assumed to be equal or greater than 1~\cite{Haug2023a}.
Moreover, since we focus on one-dimensional models away from criticality, our numerical analysis relies on tensor-network methods based on matrix product states (MPS). 
Within this framework, ground states were obtained using the density-matrix renormalization group (DMRG) algorithm~\cite{Schollwock2010, Orus2014}, and their SRE was then evaluated using two complementary approaches~\cite{Tarabunga2024, Kozic2025}. 
For ground-state determination, a modest bond dimension ($\chi \approx 50$) was sufficient, while the MPS in the Pauli basis used for the SRE evaluation was truncated at $\chi \approx 200$ - $450$ using standard variational compression techniques~\cite{Paeckel2019}.

We begin our analysis with the dimerized XX model, which can be obtained from the SSH model via a Jordan-Wigner transformation. 
Both models serve as paradigmatic examples of one-dimensional systems exhibiting SPTPs~\cite{Su1979, Batra2020}. 
Under open boundary conditions (OBCs), the Hamiltonian of the dimerized XX model takes the form 
\begin{equation}
\label{eqn:Hamiltonian_SSH}
H_{DXX} = \sum_{k=1}^{L-1} \frac{1 + (-1)^k\delta}{2} \left( \sigma^x_k \sigma^x_{k+1} + \sigma^y_{k} \sigma^y_{k+1} \right),
\end{equation}
where $L$ denotes the even length of the spin chain and $\delta \in [-1, 1]$ controls the dimerization. 
For $\delta \in [-1, 0)$, the system lies in a topologically trivial phase, while for $\delta \in (0, 1]$, it enters a topologically non-trivial phase~\cite{Asboth2016}, characterized by the presence of localized edge states and long-range entanglement between the chain endpoints~\cite{Campos2007}.

Motivated by the considerations made in the introduction, we expect the ground states in the topologically ordered phase to exhibit greater quantum magic than those in the trivial phase. 
In particular, for a fixed value $\bar{\delta} > 0$, we expect the magic $\mathcal{M}_2$ of the ground state at $\delta = \bar{\delta}$ to exceed that of the corresponding ground state at $\delta = -\bar{\delta}$. 
To test this hypothesis, we computed $\mathcal{M}_2$ as a function of $\delta$ for various system sizes $L$, with the results displayed in Fig.~\ref{Magic_SSH}.

From the figure, it is immediately evident that quantum magic exhibits a pronounced peak within the topologically ordered phase, which grows linearly in size, shifting toward the expected critical point $\delta = 0$ as $L$ increases.
Similar behaviors are well known in one-dimensional quantum systems. 
For instance, a closely analogous trend has been reported in the derivative of the concurrence for the transverse-field Ising model~\cite{Osterloh2002}.

\begin{figure}
\centering
\includegraphics[width=0.95\columnwidth]{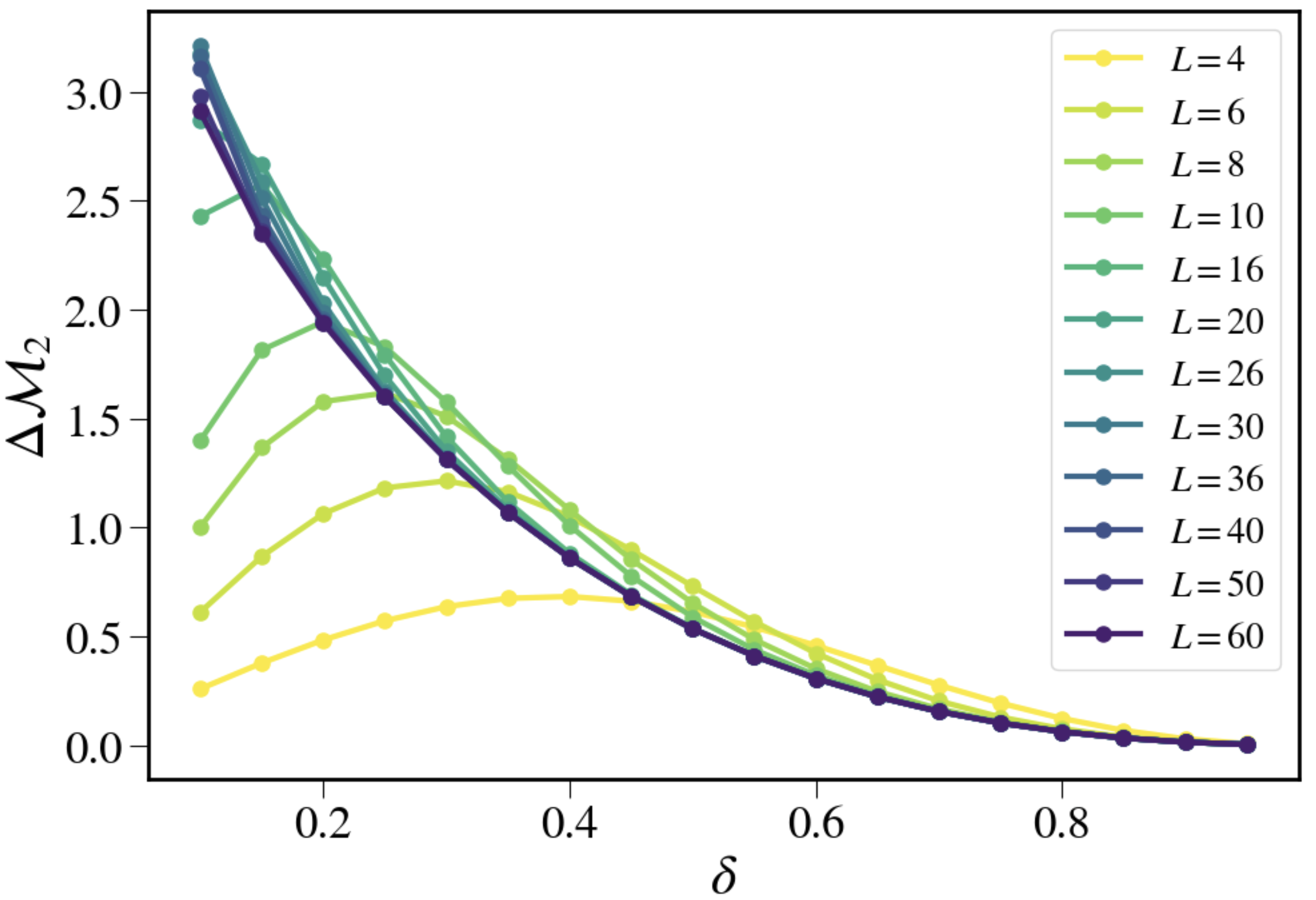}
\caption{Behavior of $\Delta \mathcal{M}_2(\delta) = \mathcal{M}_2(\delta) - \mathcal{M}_2(-\delta)$, as function of $\delta>0$ for the model in Eq.~\eqref{eqn:Hamiltonian_SSH}. 
For small $L$, $\Delta \mathcal{M}_2$ shows a size dependence, but as $L$ increases, the curves start to converge to a non-constant limiting function of $\delta$. 
This finite, parameter-dependent asymmetry is inconsistent with the behavior of a genuine topological invariant. 
The self-dual point $\delta=0$ is excluded, since $\Delta \mathcal{M}_2=0$ by definition.}
\label{DM_SSH}
\end{figure}

To isolate the potential topological contribution to SRE, for each $L$, we define the difference in magic between the ground states at symmetric points with respect to the critical point
$\Delta \mathcal{M}_2(\delta) = \mathcal{M}_2(\delta) - \mathcal{M}_2(-\delta).
$
These differences are shown in Fig.~\ref{DM_SSH}.
For $\delta>0$ $\Delta \mathcal{M}_2(\delta)$ remains consistently positive and rapidly converges to a finite value as the system size increases. 
For $L \geq 30$, the curves start to collapse onto a single well-defined limiting function except close to $\delta=0$, as the correlation length becomes comparable with the size of the system.
However, this limiting function retains a non-trivial dependence on $\delta$, suggesting that the asymmetry in quantum magic does not have a topological origin. 
In fact, topological quantities are expected to be quantized and constant within a given macroscopic phase~\cite{Micallo2020, Hamma2005, Halasz2012, Torre2024, Odavic2023}, whereas our results feature a smooth dependence on the microscopical parameters.

This suggests that the observed difference in quantum complexity is not due to topological order, but rather originates from the violation of duality symmetry induced by OBCs. 
To check this hypothesis, we repeated our evaluation assuming periodic boundary conditions (PBCs), which can be implemented by adding the term $\tfrac{1 + \delta}{2},(\sigma^x_1 \sigma^x_L + \sigma^y_1 \sigma^y_L)$ to the Hamiltonian in Eq.~\eqref{eqn:Hamiltonian_SSH}. 
Under these conditions, the dimerized XX model retains an exact duality symmetry around $\delta=0$~\cite{Asboth2016}, since the cases $\delta>0$ and $\delta<0$ are related by a one-site translation. Consequently, Pauli strings are biunivocally mapped into Pauli strings, and it can be proven (and we numerically checked), that under PBCs the SRE remains symmetric between the two phases, i.e., $\Delta \mathcal{M}_2(\delta)=0$ for all $\delta$.
The non-constant asymmetry observed with OBCs, and its absence with PBCs, therefore demonstrates that the increased complexity arises from a violation of duality symmetry rather than from topological order.
\begin{figure}
\centering
\includegraphics[width=0.95\columnwidth]{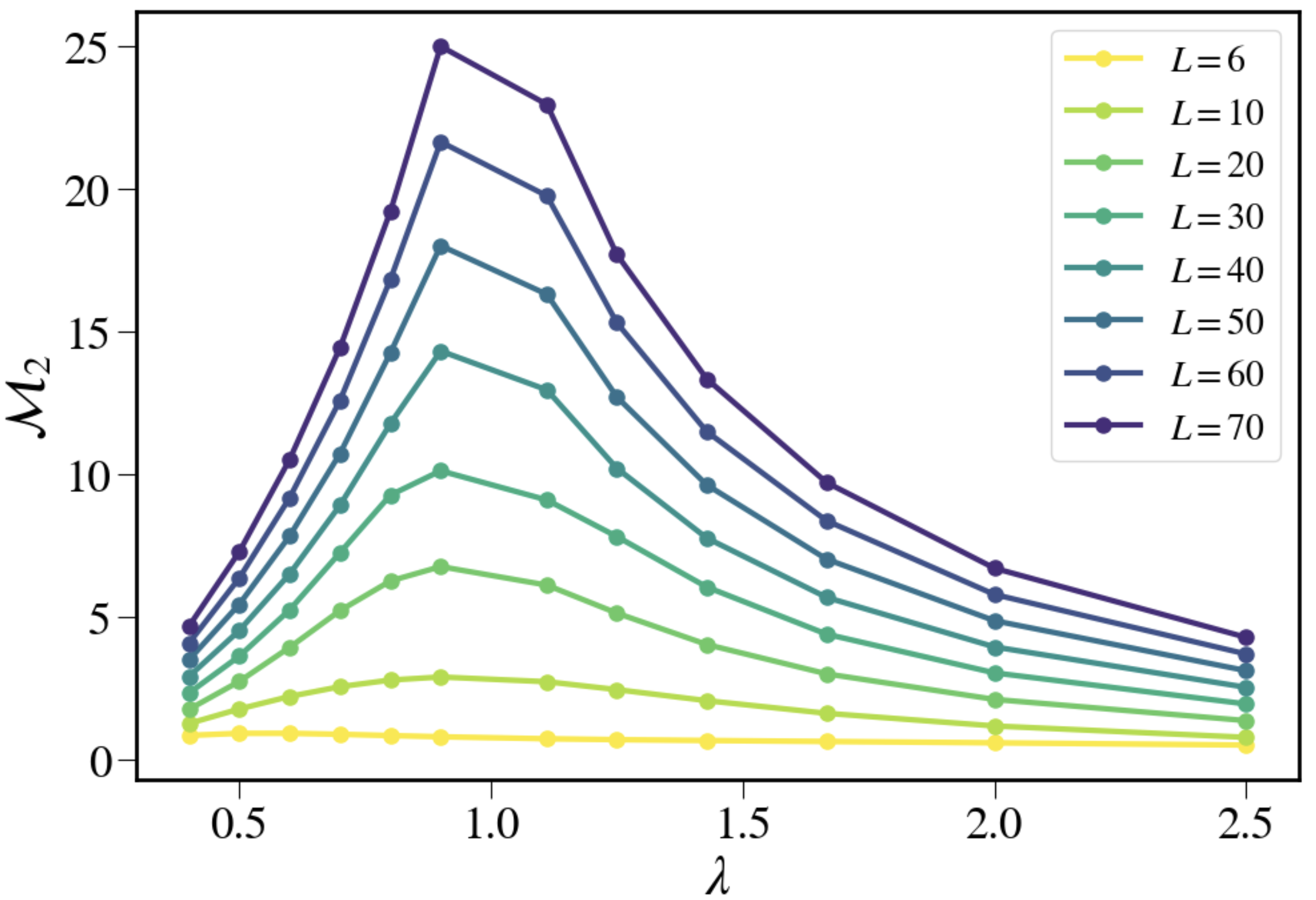}
\caption{Behavior of the magic $\mathcal{M}_2$ for the Cluster Ising model in Eq.~\eqref{eqn:Hamiltonian_CI} as a function of the parameter $\lambda$.}
\label{Magic_CI}
\end{figure}

To assess whether the observed phenomenon is specific to the dimerized XX model or indicative of a broader trend among one-dimensional spin systems hosting SPTPs, we extended our analysis to the so-called cluster-Ising model. 
The cluster-Ising model~\cite{Smacchia2011, Son2011, Doherty2009} is a one-dimensional spin-$1/2$ system whose Hamiltonian, assuming OBCs, is given by
\begin{equation}
\label{eqn:Hamiltonian_CI}
H_{\text{CI}}= -\sum_{i=1}^{L-2} \sigma_{i}^x \sigma_{i+1}^z \sigma_{i+2}^x + \lambda \sum_{i=1}^{L-1} \sigma_{i}^y \sigma_{i+1}^y.
\end{equation}
In the thermodynamic limit, the model hosts an SPTP -- known as the cluster phase -- when the three-body cluster interaction dominates over the Ising coupling. 
In this regime, edge states appear at the boundaries. 
Conversely, for $\lambda > 1$, where the Ising term prevails, the system enters a topologically trivial antiferromagnetic phase. Similarly to the SSH model, also the Cluster Ising chain is known to possess a duality symmetry, realized between the points $\lambda$ and $1/ \lambda$~\cite{Smacchia2011}.

\begin{figure}
\centering
\includegraphics[width=0.95\columnwidth]{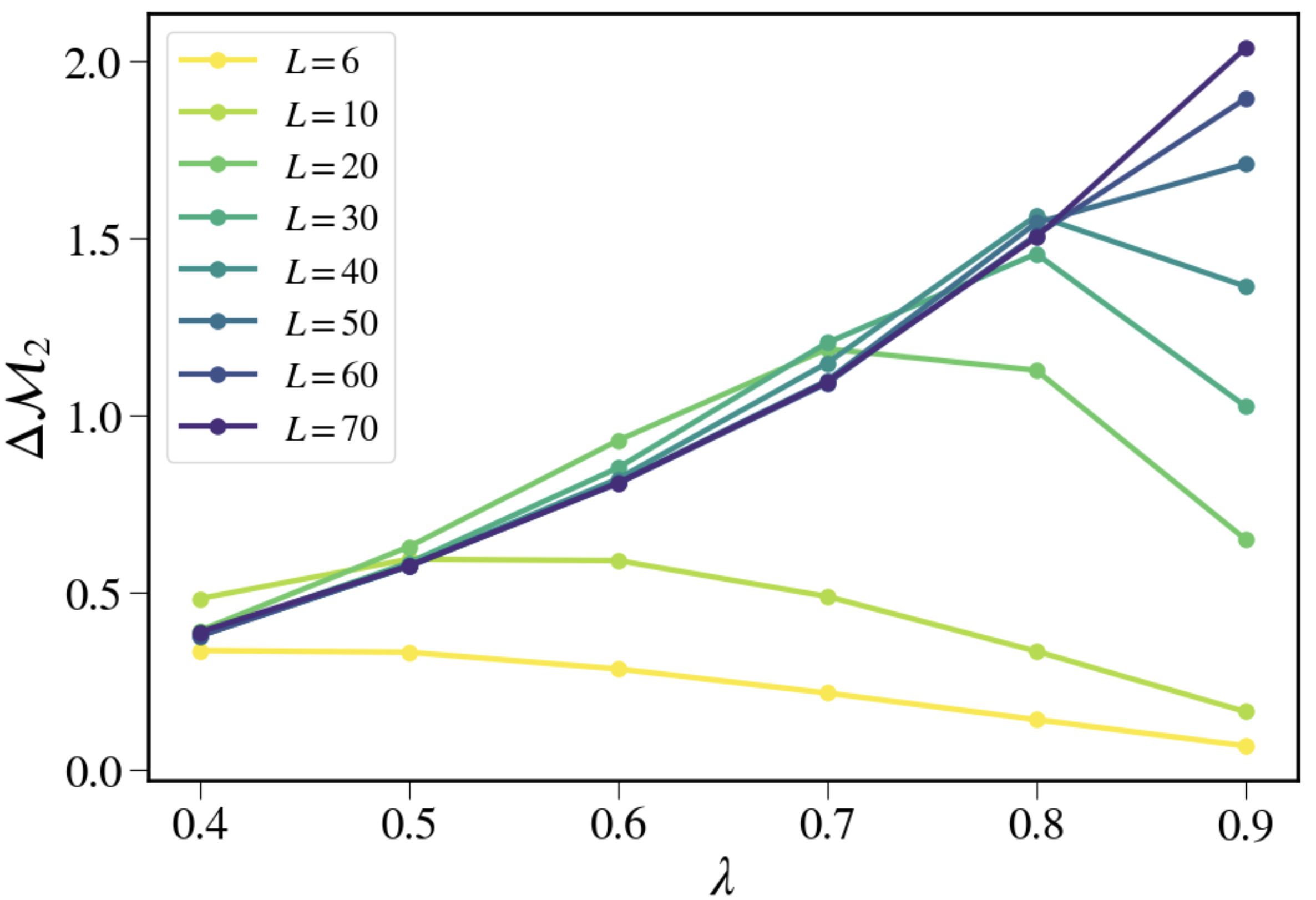}
\caption{Behavior of the magic difference between dual points \mbox{$\lambda \leftrightarrow 1/\lambda$} for the cluster-Ising model in Eq.~\eqref{eqn:Hamiltonian_CI} as a function of $\lambda$, with OBCs. For large system sizes $L$, the curve converges to a finite, $\lambda$-dependent value, indicating a non-topological origin.}
\label{DM_CI_OBC}
\end{figure}

As shown in Fig.~\ref{Magic_CI}, this system exhibits a qualitatively similar trend to that observed in the dimerized XX/SSH model. 
By defining the quantity $\Delta \mathcal{M}_2(\lambda) = \mathcal{M}_2(\lambda) - \mathcal{M}_2(1/\lambda)$ -- which again measures the difference in quantum magic between ground states at dual points -- we observe in Fig.~\ref{DM_CI_OBC} that for all $\lambda \in [0,1)$, this difference rapidly converges to a limiting curve as the system size increases. 
However, this limiting function retains a smooth, non-universal dependence on $\lambda$, precluding an interpretation in terms of a topological invariant.

Furthermore, consistent with the findings for the dimerized XX model, when the boundary conditions are changed from open to periodic -- closing the loop by adding to the Hamiltonian in Eq.~\eqref{eqn:Hamiltonian_CI} the terms $ -\sigma_{L-1}^x \sigma_{L}^z \sigma_{1}^x -\sigma_{L}^x \sigma_{1}^z \sigma_{2}^x + \lambda \sigma_1^y \sigma_L^y$ -- the quantity $\Delta \mathcal{M}_2(\lambda)$ vanishes identically. 
PBCs restore the exact duality $\lambda \leftrightarrow 1/\lambda$ and remove the possibility of observing edge states, but they do not destroy the topological nature of the $\lambda<1$ phase, which remains signaled by a finite string-order parameter~\cite{Smacchia2011}. 
Thus, exactly dual points have the same SRE (since, once again, the duality transformation is of the Clifford type), demonstrating the inability of quantum magic to detect the presence of topological order. 
Once again, the difference in SRE arises solely from the violation of duality symmetry at the boundaries. We remark that this is in line with the results of~\cite{wei2025}, where it was proven that in 1D cluster states have high computational complexity from the point of view of a Montecarlo simulation (due to their sign problem structure), but not from a Clifford resource point of view.

As a final check, let us now consider a model with an exact duality between two phases but without topological order: the transverse-field Ising chain. Its Hamiltonian with OBCs is
\begin{equation}
\label{eq:Hamiltoniam_Ising}
H_{TI} = - \sum_{i=1}^{L-1} \sigma_{i}^z \sigma_{i+1}^z - h \sum_{i=1}^L \sigma_i^x.
\end{equation}
This model undergoes a quantum phase transition at $h=1$, separating a ferromagnetic phase ($h < 1$) from a paramagnetic phase ($h >1$)~\cite{Barouch1971, FranchiniBook}. 
Crucially, although edge modes can be isolated in the ferromagnetic phase, the asymptotic degeneracy between the two lowest-energy states makes their long-range correlations fragile, and therefore the Ising chain does not host an SPTP~\cite{Pfeuty1970, Kitaev_2001}. 
This model is also well known for its self-duality $h \leftrightarrow 1/h$ around the critical point $h=1$, via the Kramers–Wannier transformation~\cite{Kramers1941}. However, the duality is exact only assuming PBCs, obtained by adding the term $-\sigma_1^z \sigma_L^z$ to Eq.~\eqref{eq:Hamiltoniam_Ising}.

\begin{figure}
\centering
\includegraphics[width=0.95\columnwidth]{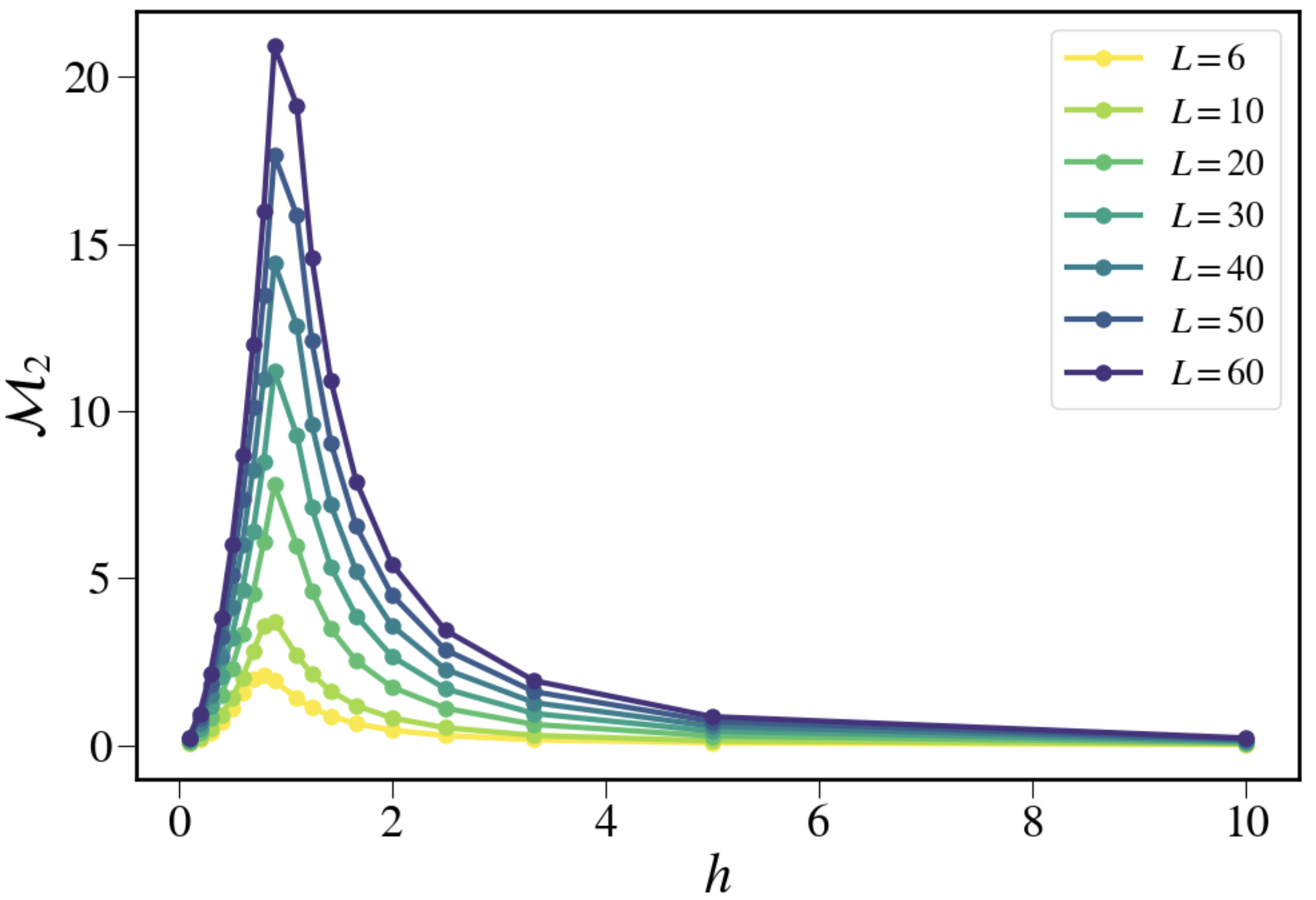}
\caption{Behavior of $\mathcal{M}_2$ for the Ising model in eq.~\eqref{eq:Hamiltoniam_Ising}, as a function of the transverse magnetic field $h$, with OBCs. 
Although this model does not support topological order, quantum magic behaves surprisingly similarly to the topological model considered before.}
\label{M_Ising}
\end{figure}

\begin{figure}
\centering
\includegraphics[width=0.95\columnwidth]{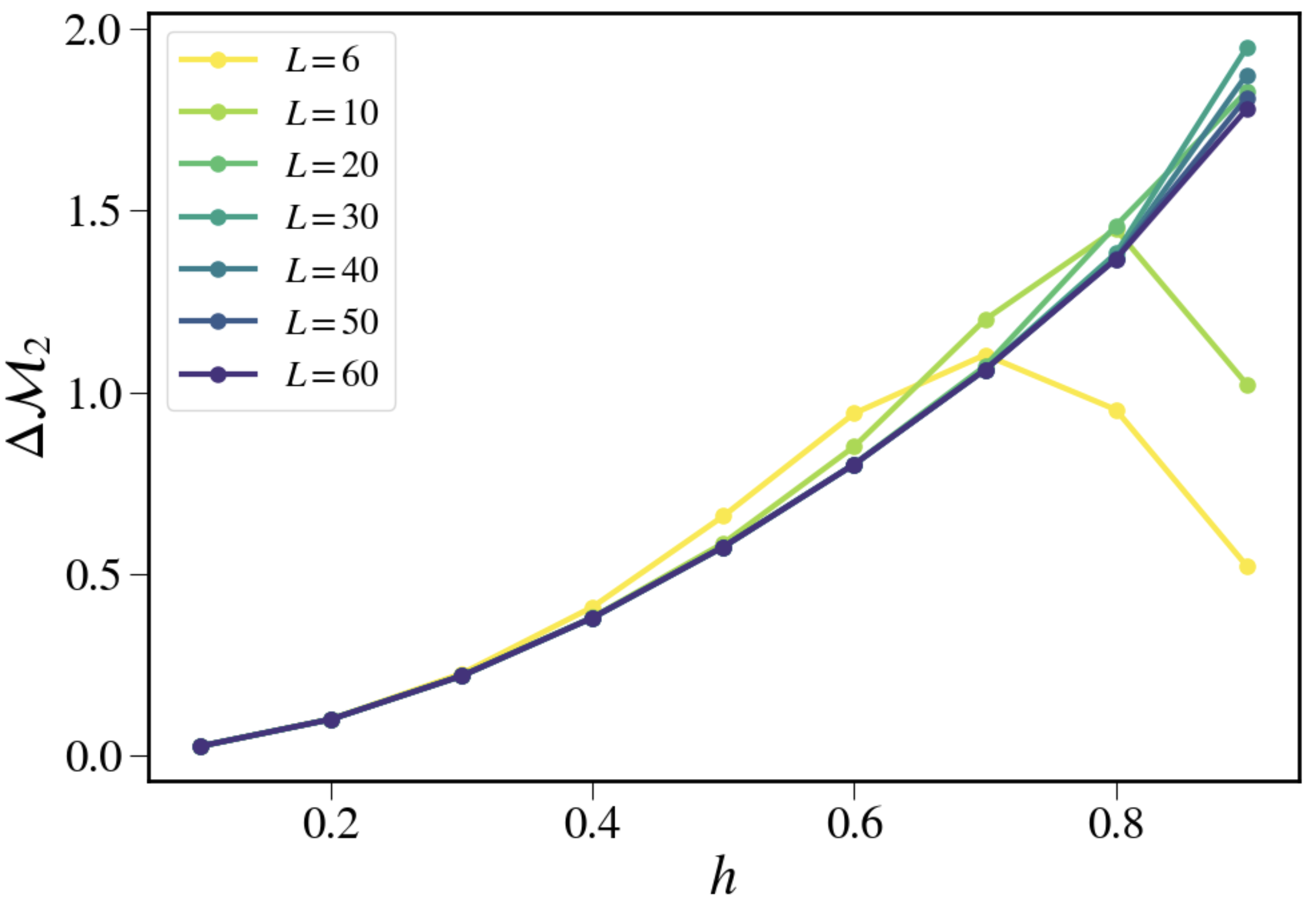}
\caption{Difference $\Delta \mathcal{M}_2$ between SRE calculated at mirror point $h \longleftrightarrow 1/h$ for the Ising chain in eq.~\eqref{eq:Hamiltoniam_Ising} as a function of the external magnetic field $h$, with OBCs. As observed for the topological models, $\Delta \mathcal{M}_2$ is finite and non-constant, growing approaching the critical point $h=1$, although this model does not support topological order.}
\label{DM_Ising_OBC}
\end{figure}

As can be appreciated from Fig.~\ref{M_Ising}, the behavior of quantum magic $\mathcal{M}_2$ for the ground state of the transverse field Ising model is qualitatively the same as that of the previous models: with OBCs, $\mathcal{M}_2$ has a peak inside the ferromagnetic phase, which grows and moves towards $h=1$ as the system size is increased. 
This behavior is sensibly different from what was obtained in~\cite{Oliviero2022}, where the same model was analyzed assuming PBCs, i.e., with exact dual symmetry, and agrees with what we get under the same conditions: the maximum is always centered at $h=1$.
Moreover, the quantity $\Delta \mathcal{M}_2(h) = \mathcal{M}_2(h) - \mathcal{M}_2(1/h)$ ($h\in[0,1)$) vanishes identically when the duality symmetry holds, independently of $h$ and $L$ because the Kramers-Wannier duality is magic-preserving. 
Instead, as also observed for the other models, breaking the duality induces an asymmetry in the SRE. 
As shown in Fig.~\ref{DM_Ising_OBC}, with OBCs $\Delta \mathcal{M}_2(h)$ vanishes only in the limit of $h\rightarrow 0$. Otherwise, its value depends on $L$, but the different curves tend to collapse as soon as the dimension of the system becomes greater than the correlation length.  
Obviously, since no topological phase is present in this model, the existence of such a difference cannot be justified by the presence of topological order.  

\begin{table}[b]
\centering
\begin{tabular}{|c|c|c|c|c|c|}
\hline
\multirow{ 2}{*}{Model} & Ham. & Boundary  & 1 SPTP & Exact & SRE \\ 
& Eq. & Conditions & phase & Duality & Asymmetry  \\ \hline
Dimerized XX/ & \multirow{ 2}{*}{\eqref{eqn:Hamiltonian_SSH}} & OBCs & \ding{51} & \ding{55} & \ding{51} \\ SSH && PBCs & \ding{51} & \ding{51} & \ding{55}  \\ \hline
\multirow{ 2}{*}{Cluster Ising} & \multirow{ 2}{*}{\eqref{eqn:Hamiltonian_CI}} & OBCs & \ding{51} & \ding{55} & \ding{51} \\ && PBCs & \ding{51} & \ding{51} & \ding{55}  \\ \hline
\multirow{ 2}{*}{Ising Chain} & \multirow{ 2}{*}{\eqref{eq:Hamiltoniam_Ising}} & OBCs & \ding{55} & \ding{55} & \ding{51} \\ && PBCs & \ding{55} & \ding{51} & \ding{55}  \\ \hline
\end{tabular}
\caption{Recap of our results: for every model we considered we report whether, with Open or Periodic Boundary Conditions (OBCs and PBCs respectively), it hosts a SPTP or not, if the duality symmetry is exact, and if the SRE is the same between dual points or asymmetric.}
\label{Table:recap}
\end{table}

Before concluding, we emphasize that all evaluations of quantum magic in this Letter were carried out in the computational basis.
Since magic is inherently basis-dependent, one may wonder whether our results are tied to this choice or hold more generally. 
Given the difficulty of evaluating magic, a comprehensive analysis over all possible bases in all possible systems is not feasible.
Nevertheless, by restricting our study to small systems ($L \leq 10$) and to cases where the change of basis is implemented through unitaries composed of independent single-spin rotations, such an analysis becomes tractable.
Our framework demonstrates notable robustness: under periodic boundary conditions (PBCs), the magic at the dual points remains identical across all bases, whereas opening the chain—thus breaking duality and inducing the emergence of edge states—breaks this symmetry.
Hence, the physical conclusions presented here are not restricted to a specific basis but acquire a general validity. 
Finally, from our results, we recover that the computational basis minimizes the magic for all models and parameter choices, and thus coincides with the non-local $L$-partite magic~\cite{Cao2023, Cao2025, Qian2025}.


In conclusion, we have analyzed the rank-$2$ Stabilizer R\'enyi Entropy $\mathcal{M}_2$ to investigate the possible contribution of topological order to quantum magic in 1D SPTPs. 
To isolate it, we focused on models exhibiting an exact duality, namely the dimerized XX model and the Cluster–Ising model. 
Since SPTPs are generally regarded as valuable quantum resources expected to yield computational advantage within the MBQC framework, an enhancement of $\mathcal{M}_2$ between dual points might be anticipated. 
On the contrary, we find that the dual phases exhibit identical amounts of magic. 
A subextensive asymmetry appears only under open boundary conditions, where the duality is broken by edge effects, introducing finite corrections to $\mathcal{M}_2$. 
This difference is not constant across the phase and, as confirmed by the identical behavior observed in the quantum Ising chain, is of non-topological origin. 
These results are summarized in Table~\ref{Table:recap}. 
Our findings stand in stark contrast to the case of topological frustration in one dimension~\cite{Maric2020b, MaricFate, Torre2024, koziccoherence2025}, where the introduction of a delocalized topological excitation into the ground state leads to an additional contribution to quantum magic that grows logarithmically with system size~\cite{Odavic2023, Catalano2025}.

Since $\mathcal{M}_2$ quantifies the amount of non-Clifford resources required to generate a given state, our results indicate that the ground states of SPTPs can be efficiently prepared using the same resources as their topologically trivial counterparts. 
However, given that these phases differ fundamentally within the MBQC paradigm, this observation appears to challenge the expected equivalence of computational complexity across different quantum computation models and thus calls for a deeper understanding. 
On the other hand, our findings are consistent with recent results showing that Local Operations and Classical Communication (LOCC) can convert local into global entanglement~\cite{Piroli2021, Piroli2024, Lu2022, Tantivasadakarn2023, Smith2023, Smith2024, Sahay2025}, suggesting that even intrinsic topological order may not be a necessary resource in non-unitary computational schemes. 
Nevertheless, the classification of these schemes within a Clifford resource framework remains nontrivial, and it is therefore unclear whether our results are a manifestation of such mechanisms or evidence of a distinct underlying phenomenon.

To clarify the generality of our findings they should be extended to truly topological models, with both abelian and non-abelian symmetries, in higher dimensions: in absence of exact dualities, one could check whether the growth of $\mathcal{M}_2$ with the system size deviates from the one expected considering purely local order~\cite{Leone2022} to assess whether topological order carries or not an intrinsic amount of quantum complexity. 

\begin{acknowledgments}
The research leading to these results has received funding from the following organizations: European Union via ICSC - Italian Research Center on HPC, Big Data and Quantum Computing (NextGenerationEU Project No. CN00000013), project EuRyQa (Horizon 2020), project PASQuanS2 (Quantum Technology Flagship); Italian Ministry of University and Research (MUR) via: Quantum Frontiers (the Departments of Excellence 2023-2027); the World Class Research Infrastructure - Quantum Computing and Simulation Center (QCSC) of Padova University; Istituto Nazionale di Fisica Nucleare (INFN): iniziativa specifica IS-QUANTUM.
A.G.C. acknowledges support from the MOQS ITN programme, a European Union’s Horizon 2020 research and innovation program under the Marie Sk\l{}odowska-Curie grant agreement number 955479. 
G. T., F. F., and S. M. G. acknowledge support of ``Implementation of cutting-edge research and its application as part of the Scientific Center of Excellence for Quantum and Complex Systems, and Representations of Lie Algebras", Grant No. PK.1.1.10.0004, co-financed by the European Union through the European Regional Development Fund - Competitiveness and Cohesion Programme 2021-2027.
S. B. K. acknowledges support from the Croatian Science Foundation (HrZZ) Projects DOK-2020-01-9938. 
S. P. acknowledges financial support from the Ministero dell’Universitá e della Ricerca (MUR) and the Project PRIN 2022 number 2022W9W423 funded by the European Union Next Generation EU. 
\end{acknowledgments}

\bibliography{biblio-01}

\end{document}